\documentclass[english,prl, showpacs, twocolumn]{revtex4}
\usepackage[latin1]{inputenc}
\usepackage{graphicx}
\usepackage{amssymb}

\makeatletter

\usepackage{babel}
\makeatother
\begin{document}

\title{Dynamical multistability in high-finesse micromechanical optical
cavities}

\author{Florian Marquardt, J. G. E. Harris and S. M. Girvin }

\affiliation{Departments of Physics and Applied Physics, Yale University, PO Box
208284, New Haven, CT 06520 (USA)}

\begin{abstract}
We analyze the nonlinear dynamics of a high-finesse optical cavity
in which one mirror is mounted on a flexible mechanical element. We
find that this system is governed by an array of dynamical attractors,
which arise from phase-locking between the mechanical oscillations
of the mirror and the ringing of the light intensity in the cavity.
We describe an analytical approximation to map out the diagram of
attractors in parameter space, derive the slow amplitude dynamics
of the system, including thermally activated hopping between different
attractors, and suggest a scheme for exploiting the dynamical multistability
in the measurement of small displacements.
\end{abstract}

\pacs{07.10.Cm, 42.79.Gn, 05.45.-a}

\maketitle
\emph{Introduction}. - The radiation pressure exerted by light stored
in an optical cavity couples the mechanical degrees of freedom of
the cavity mirrors to the optical field. The dynamics which result
from this coupling offer novel means for manipulating both the light
and the mirrors themselves, for example by generating squeezed \cite{1994_02_Fabre_SqueezingInCavityWithMovableMirror}
or entangled states of light \cite{2001_06_Giovannetti_RadiationPressureInducedEPR}
or matter \cite{2003_09_Marshall_QSuperposMirror}, or by tailoring
the mechanical properties of the mirrors \cite{1967_BraginskyManukin_PonderomotiveEffectsEMRadiation,1983_10_DorselWalther_BistabilityMirror,1985_11_Meystre_RadiationPressureDrivenInterferometers,2003_08_Vogel_PhotothermalForceOnCantilever,2004_12_HoehbergerKarrai_CoolingMicroleverNature}.
The radiation pressure also sets fundamental and practical limits
on the sensitivity of a range of experiments, from the km-scale optical
cavities in gravitational wave observatories such as LIGO to the $\mu{\rm m}$-scale
cavities used to probe the motion of micromechanical systems, e.g.
by enforcing the standard quantum limit for displacement measurements
\cite{1980_07_Caves_ShotNoiseLimitedInterferometricLengthMeasurement,2003_04_Braginsky_InterferometerNoiseIndependentOfTestMassQuantization,1999_02_Tittonen_InterferometerMeasurementMacroBody,1995_01_Sidles_RMP_MagneticForceMicroscopy}
or introducing mechanical instabilities \cite{1987_10_AguirregabiriaBel_DelayInducedInstabilityFP,1988_03_Bel_DynamicsPendularFabryPerotCavity,1994_02_Fabre_SqueezingInCavityWithMovableMirror,2001_07_Braginsky_ParametricInstabilityFPCavity}.

To date the theoretical description of these systems has been almost
entirely in terms of linearized equations of motion, valid for small
mirror oscillations. While this approach has been adequate to predict
a wide range of novel effects, it neglects the inherent nonlinearity
of these devices and cannot describe their behavior in dynamically
unstable regimes. As advances in micromachining \cite{1998_03_ClelandRoukes_NanometreScaleMechanicalElectrometer,1999_08_Harris_CantileverMagnetometry,2001_05_Harris_Magnetization2DEG_PRL,2004_07_Rugar_SingleSpinMRFM}
and optical cavity fabrication \cite{1992_03_Rempe_UltralowLossInterferometer,2001_08_Hood_HighFinesseMirrors}
produce devices with stronger optomechanical coupling and weaker damping
than was previously possible, the nonlinearities and dynamical instabilities
of these systems will become increasingly important. Here we present
an analytic theory which provides a full description of the nonlinear
regime. We find that the dynamics is dominated by an array of stable
dynamical attractors which correspond to phase-locking between the
mirror motion and the ringing in the cavity optical field. These results
are confirmed in detail by numerical simulations of the microscopic
equations of motion. The theory presented here also describes the
statistical behavior of the cantilever oscillation amplitude and thermally-activated
transitions between the dynamical attractors. 

\begin{figure}
\includegraphics[%
  width=1.0\columnwidth]{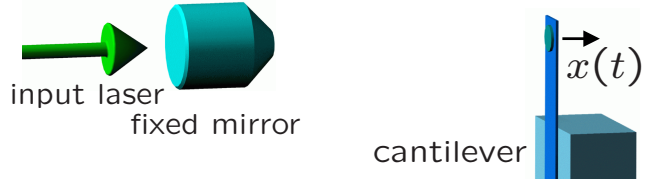}

\caption{\label{cap:The-setup-analyzed}The setup analyzed in the text (not
drawn to scale).}
\end{figure}

\emph{The model}. - We consider the setup of Fig. \ref{cap:The-setup-analyzed}.
The coupling between the cantilever position and the light intensity
of a given cavity mode is described by the Hamiltonian $\hat{H}_{{\rm cav}}=\hbar\omega_{L}(1-\hat{x}/l)\hat{a}^{\dagger}\hat{a}$,
provided we may neglect the finite travel-time of the light between
the mirrors. The cavity of length $l$ is resonant with the incoming
laser radiation of frequency $\omega_{L}$ when $x=0$, and we assume
the excursions in $x$ to be small enough to avoid the other resonances.
This coupling gives rise to a radiation pressure force $\hat{F}=(\hbar\omega_{L}/l)\hat{a}^{\dagger}\hat{a}$.
In the following we will consider the purely classical nonlinear dynamics,
where $\hat{a}$ is replaced by the coherent light amplitude $\alpha$.
We use the ring-down time of the cavity $\gamma^{-1}$, the resonance
width $\delta x=l\gamma/\omega_{L}$, and the cantilever mass $m$
as convenient new units, and rescale $\alpha$ such that it becomes
$1$ at resonance: $\alpha=\alpha^{{\rm orig}}e^{i\omega_{L}t}/\sqrt{n_{{\rm max}}}$,
where the maximum photon number $n_{{\rm max}}=4P_{{\rm in}}/(\gamma\hbar\omega_{L})$
is linear in the input power $P_{{\rm in}}$. Then the coupled equations
of motion for $\alpha$ and $x$ read \cite{1994_02_Fabre_SqueezingInCavityWithMovableMirror}:

\begin{eqnarray}
\dot{\alpha} & = & \left(ix-\frac{1}{2}\right)\alpha+\frac{1}{2}\label{alphaEq}\\
\ddot{x} & = & \mathcal{P}|\alpha|^{2}-\omega_{0}^{2}(x-x_{0})-\Gamma\dot{x}\label{Xeq}\end{eqnarray}
The oscillator frequency $\omega_{0}$ and mechanical damping rate
$\Gamma$ are fixed, while the detuning from resonance $x_{0}$ may
be controlled, either by changing the laser frequency or applying
a static force to the cantilever. All the other constants are combined
into the dimensionless input power

\begin{equation}
\mathcal{P}=\frac{4P_{{\rm in}}\omega_{L}}{m\gamma^{4}l^{2}}=\frac{\hbar n_{{\rm max}}}{m\delta x^{2}\gamma}\,.\end{equation}
 We emphasize that the scale of the radiation pressure, set by $\mathcal{P}$,
grows with the fourth power of the cavity finesse ($\mathcal{P}\propto\gamma^{-4}$
with $\gamma=(c/2l)\mathcal{T}$, where $\mathcal{T}$ is the mirror
transmission). The nonlinear effects we investigate become important
for $\Gamma\ll\mathcal{P}$.

This work is motivated in part by the possibility of using the fabrication
process described in \cite{1999_08_Harris_CantileverMagnetometry,2001_05_Harris_Magnetization2DEG_PRL}
to integrate ultrasensitive cantilevers with high-reflectivity dielectric
mirrors, and to use these mirrors as part of a high-finesse optical
cavity at low temperatures. In \cite{2001_05_Harris_Magnetization2DEG_PRL},
cantilevers with spring constants of $\sim10^{-3}{\rm N/m}$ and mass
$10^{-11}{\rm kg}$ were integrated with lithographically-patterned
high-quality heterostructures similar in design to high-reflectivity
dielectric mirrors. These cantilevers were then used in an optical
interferometer with an incident power $P_{{\rm in}}\sim20{\rm nW}$
at $T\sim0.3\, K$. By making modest adjustments to the dimensions
of these devices, it should be possible to realize cantilevers with
spring constant $\sim1\,{\rm N/m}$, mass $m\sim10^{-10}{\rm kg}$,
and mechanical $Q\sim10^{5}$ supporting a mirror capable of achieving
a cavity finesse of $\pi/\mathcal{T}\sim10^{5}$, where $l\sim10{\rm cm}$.
Assuming these values, the cavity ringdown rate $\gamma\sim10^{5}{\rm Hz}$
becomes comparable with the cantilever frequency ($\omega_{0}\sim1$
in units of $\gamma$), and the dimensionless damping rate $\Gamma\sim10^{-5}$
is much smaller than $\mathcal{P}\sim1$. In contrast, our estimate
for the recent low-finesse experiment on radiation cooling of a cantilever
\cite{2004_12_HoehbergerKarrai_CoolingMicroleverNature} is roughly
$\mathcal{P}\sim10^{-18}\ll\Gamma$ (both for radiation-pressure and
photothermal effects). 

The cavity resonance peak $\alpha(x)=1/(1-2ix)$ gives rise to a rounded
step-like barrier in the effective static cantilever potential obtained
by integrating the right-hand-side (rhs) of Eq. (\ref{Xeq}),

\begin{equation}
V_{{\rm eff}}(x)=\frac{\omega_{0}^{2}}{2}(x-x_{0})^{2}-\frac{\mathcal{P}}{2}\arctan(2x)\,.\label{Veff}\end{equation}
Depending on the parameters $(\omega_{0},x_{0},\mathcal{P})$, there
are one or two local minima of $V_{{\rm eff}}$, opening the possibility
of \emph{static} bistability \cite{1983_10_DorselWalther_BistabilityMirror,1985_11_Meystre_RadiationPressureDrivenInterferometers}.
However, it is known \cite{1967_BraginskyManukin_PonderomotiveEffectsEMRadiation}
that the time-lag of the radiation pressure force generated by the
finite cavity ring-down time $\gamma^{-1}$ introduces additional
damping or anti-damping to the left or right of the barrier, respectively.
Here we will focus on the regime where the anti-damping leads to a
linear dynamical instability discovered previously \cite{1987_10_AguirregabiriaBel_DelayInducedInstabilityFP,1988_03_Bel_DynamicsPendularFabryPerotCavity,1994_02_Fabre_SqueezingInCavityWithMovableMirror,2000_02_Pai_RadiationPressureInstabilityLIGO},
such that there is no stable stationary solution $\dot{x}=0$. Then
the system settles into self-sustained oscillations, whose full nonlinear
dynamics we explore here. 

\begin{figure}
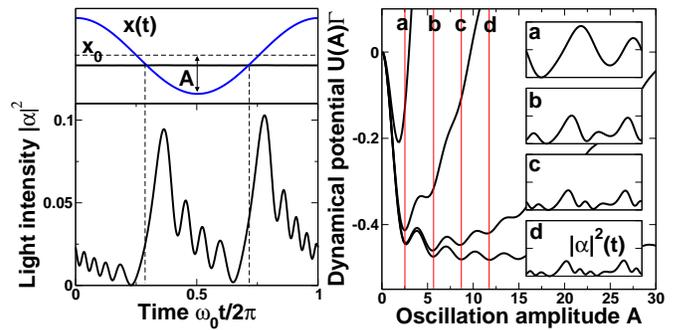

\includegraphics[%
  width=0.48\columnwidth]{AlphaSquare.eps}~\includegraphics[%
  width=0.52\columnwidth]{Potential.eps}

\caption{\label{cap:Light-intensity-oscillations}Left: Light intensity oscillations
for given sinusoidal cantilever motion (with $\omega_{0}=1$, $\bar{x}=x_{0}=5,\, A=20$).
Right: Resulting dynamical potential $U(A)$, for $\mathcal{P}=1$
and $\Gamma=10^{-4},10^{-3},10^{-2},10^{-1}$ (bottom to top curve).
Insets show $|\alpha|^{2}(t)$, with time on the horizontal axis spanning
one period and the vertical scale equal to $0.4$ in each case. Note
the additional oscillation in $|\alpha|^{2}$ for each subsequent
potential well, i.e. each new dynamical attractor.}
\end{figure}
\emph{Dynamics in the unstable regime}. - Direct numerical solution
of Eqs. (\ref{alphaEq}), (\ref{Xeq}) reveals that $x(t)$ carries
out approximately sinusoidal oscillations at the unperturbed frequency
$\omega_{0}$, 

\begin{equation}
x(t)\approx\bar{x}+A\cos(\omega_{0}t)\,.\label{xOft}\end{equation}
This fact is essential in allowing us to derive an analytical theory
for the nonlinear cantilever dynamics. The light amplitude $\alpha(t)$
develops a more complicated behaviour (Fig. \ref{cap:Light-intensity-oscillations}).
It experiences a sharp rise each time the cantilever swings through
the resonance at $x=0$. Its dynamics resembles that of a driven damped
oscillator which is swept through resonance non-adiabatically. The
exact solution for a given $x(t)$ (Eq. (\ref{xOft})) can be written
as a Fourier series, $\alpha(t)=e^{i\varphi(t)}\sum_{n}\alpha_{n}e^{in\omega_{0}t}$,
with

\begin{equation}
\alpha_{n}=\frac{1}{2}\frac{J_{n}\left(-\frac{A}{\omega_{0}}\right)}{in\omega_{0}+\frac{1}{2}-i\bar{x}}\,,\label{alphaT}\end{equation}
and a global phase $\varphi(t)=(A/\omega_{0})\sin(\omega_{0}t)$,
where $J_{n}$ is the Bessel function of the first kind.

\emph{Dynamical multistability}. - We can determine the possible dynamical
attractors $(\bar{x},A)$ of the cantilever motion by imposing two
conditions that follow from Eq. (\ref{Xeq}) for any periodic motion.
The total time-averaged force $\left\langle \ddot{x}\right\rangle $
has to vanish, and the net power input via the radiation pressure
force must equal the power dissipated through friction, $\left\langle \ddot{x}\dot{x}\right\rangle =0$.
This translates into

\begin{eqnarray}
\mathcal{P}\left\langle |\alpha(t)|^{2}\right\rangle  & = & \omega_{0}^{2}(\bar{x}-x_{0})\label{ForceBalance}\\
P_{{\rm rad}}=\mathcal{P}\left\langle |\alpha(t)|^{2}\dot{x}\right\rangle  & = & P_{{\rm fric}}=\Gamma\left\langle \dot{x}^{2}\right\rangle \,.\label{PowerBalance}\end{eqnarray}
The exact dependence on $\bar{x},A$ and $\omega_{0}$ follows directly
from the coefficients of Eq. (\ref{alphaT}):

\begin{eqnarray}
\left\langle |\alpha|^{2}\right\rangle  & = & \sum_{n}|\alpha_{n}|^{2}\\
\tilde{P}_{{\rm rad}}=\left\langle |\alpha|^{2}\dot{x}\right\rangle  & = & A\omega_{0}{\rm Im}\sum_{n}\alpha_{n}^{*}\alpha_{n+1}\label{PtildeRad}\end{eqnarray}
The power balance equation can be recast into the form

\begin{equation}
\frac{\tilde{P}_{{\rm rad}}(\bar{x},A)}{\tilde{P}_{{\rm fric}}(A)}=\frac{\Gamma}{\mathcal{P}}\,,\label{equilbCondition}\end{equation}
with $\tilde{P}_{{\rm fric}}(A)=\omega_{0}^{2}A^{2}/2$. Stable attractors
are those where the ratio decreases for increasing $A$. 

\begin{figure}
\includegraphics[%
  width=1.0\columnwidth]{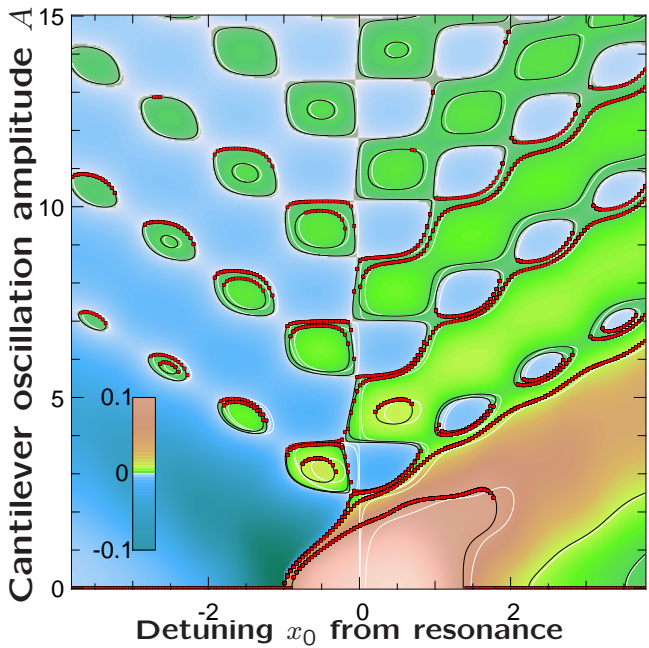}

\caption{\label{cap:Contour-lines-of}Density plot of the ratio of reduced
radiation power input and frictional power loss, $\tilde{P}_{{\rm rad}}(\bar{x}(x_{0},A),A)/\tilde{P}_{{\rm fric}}(A)$,
in the $(x_{0},A)$-plane. The contour lines indicate dynamically
stable cantilever oscillation amplitudes $A$, according to Eq. (\ref{equilbCondition}).
White contour lines display the approximation $\bar{x}\approx x_{0}$.
Contours are drawn for $\Gamma/\mathcal{P}=10^{-4},10^{-3},10^{-2},10^{-1}$
(with $\mathcal{P}=\omega_{0}=1$). Red dots show the long-time limit
($t_{{\rm sim}}=2\cdot10^{4}$) of the amplitude $A=x_{{\rm max}}-x_{{\rm min}}$
in the numerically exact solution of the original equations of motion,
obtained for the same values of $\Gamma/\mathcal{P}$ and a set of
random initial conditions.}
\end{figure}
After solving the force balance equation (\ref{ForceBalance}) for
$\bar{x}=\bar{x}(x_{0},A)$, we can plot the contour lines of the
lhs of Eq. (\ref{equilbCondition}) in the $(x_{0},A)$-plane. These
trace out the diagram of possible equilibrium values of the cantilever
oscillation amplitude $A$ as a function of detuning $x_{0}$ (Fig.
\ref{cap:Contour-lines-of}). The red dots in Fig. \ref{cap:Contour-lines-of}
show the results of a (much more time-consuming) full simulation of
the initial equations of motion, Eqs. (\ref{alphaEq}),(\ref{Xeq}),
confirming that our analytical approximation works extremely well.

The initial onset of the instability at some threshold $x_{0}^{*}<0$
(with $x_{0}^{*}=-1$ in Fig. \ref{cap:Contour-lines-of}) can be
described by expanding Eq. (\ref{ForceBalance}) for small $A$ (and
$\bar{x}\approx0$), which is valid in the limit $\Gamma\ll\mathcal{P}$.
This leads to $A\approx[(1+4\omega_{0}^{2})(1+\omega_{0}^{2}x_{0}/\mathcal{P})/2]^{1/2}$.
The position of the onset, $x_{0}^{*}=-\mathcal{P}/\omega_{0}^{2}$,
and the rise of $A$ do not depend on $\Gamma$ in this limit. 

Near the onset we have $|\bar{x}|\ll x_{0}$. However, for the rest
of parameter space $(x_{0},A)$, we find that $\bar{x}\approx x_{0}$
is a good approximation (this approximation is shown as the white
lines in Fig. \ref{cap:Contour-lines-of}, as compared to the black
lines), becoming exact for $\mathcal{P}\rightarrow0$. Within this
approximation, which we will adopt in the following, we only need
to fulfill the power balance equation (\ref{PowerBalance}). Its lhs
does not depend on the damping $\Gamma$ or the dimensionless light
power $\mathcal{P}$, and therefore the equilibrium values of $A$,
at a given $x_{0}$, depend only on the ratio $\Gamma/\mathcal{P}$
within this approximation. 

At large $A/\omega_{0}$, we find 

\begin{equation}
\tilde{P}_{{\rm rad}}\approx\frac{-\omega_{0}}{2(1+\omega_{0}^{2})}\cos(2\frac{A}{\omega_{0}}){\rm Im}\,\left[{\rm \sinh}\left(\frac{(\frac{1}{2}-ix_{0})\pi}{\omega_{0}}\right)\right]^{-1}.\label{approxPrad}\end{equation}
This reveals explicitly the periodic structure observed in Fig. \ref{cap:Contour-lines-of},
with periods of $\pi\omega_{0}$ and $2\omega_{0}$ in $A$ and $x_{0}$,
respectively. Physically, an increase in $A$ enlarges the time-dependent
detuning $x(t)$ which is equal to the frequency of the ringing in
the light intensity $|\alpha|^{2}(t)$ (Fig. \ref{cap:Light-intensity-oscillations}).
The appearance of the array of attractors is thus due to a phase-locking
phenomenon, where the oscillations of the light intensity (depending
on $A$) seek to be commensurate with the fundamental cantilever period.
For the experimental parameters mentioned above, the periods in $A$
and $x_{0}$ will be on the picometer scale.

\begin{figure}
\includegraphics[%
  width=1.0\columnwidth]{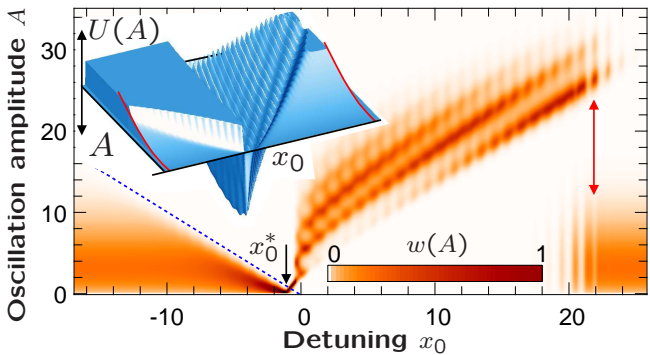}

\caption{\label{cap:BoltzmannDist}Boltzmann distribution $w(A)$ of the cantilever
oscillation amplitude $A$, as a function of detuning $x_{0}$, for
a reduced temperature $\tilde{T}=10$. Other parameters: $\mathcal{P}=\omega_{0}=1,\,\Gamma=10^{-3}$.
Note the sharp drop in $w(A)$ at $A=-x_{0}$ (dashed blue line),
the narrowing (cooling) of the distribution for $x_{0}<0$, the onset
of instability at $x_{0}^{*}$, and the transition(s) above $x_{0}=20$,
back to the thermal distribution of an unperturbed oscillator. Inset:
Underlying effective potential $U(A)$ (top truncated), with bare
oscillator parabolic potential $\omega_{0}^{2}A^{2}/2$ indicated
in red.}
\end{figure}

\emph{Amplitude equation of motion}. - We can determine the slow dynamics
of the cantilever oscillation amplitude $A$ by equating the change
in total cantilever energy $E=\frac{\omega_{0}^{2}}{2}A^{2}$ to the
net power input:

\begin{equation}
\frac{dE}{dt}=P_{{\rm total}}=P_{{\rm rad}}-P_{{\rm fric}}\label{dEdt}\end{equation}
This can be used to obtain the overdamped motion of $A$,

\begin{equation}
\frac{dA}{dt}=\frac{1}{A\omega_{0}^{2}}P_{{\rm total}}=-\frac{\Gamma}{2\omega_{0}^{2}}U'(A)\,,\label{dAdt}\end{equation}
where we have introduced the effective potential for $A$,

\begin{equation}
U(A)=\frac{\omega_{0}^{2}}{2}A^{2}-\frac{2\mathcal{P}}{\Gamma}\int_{A_{0}}^{A}\tilde{P}_{{\rm rad}}(A')\frac{dA'}{A'}\,.\label{UA}\end{equation}
The first part of $U(A)$ is the oscillator potential itself, which
drives $A$ towards zero in the absence of radiation pressure ($\mathcal{P}=0$).
The second part produces a decaying oscillating component, as follows
from Eq. (\ref{approxPrad}). The total potential $U(A)$, obtained
from $\tilde{P}_{{\rm rad}}(A)$, in general displays several local
minima, corresponding to the dynamical attractors, before the quadratic
rise of the first term takes over at large $A$ (see Fig. \ref{cap:Light-intensity-oscillations},
right). Solving Eq. (\ref{dAdt}) and comparing to ab-initio simulations
of the full dynamics yields a good agreement, which can be improved
at small $|x_{0}|$ by using $\bar{x}=\bar{x}(x_{0},A)$ from Eq.
(\ref{ForceBalance}) instead of $\bar{x}=x_{0}$. This was also used
for Fig. \ref{cap:BoltzmannDist}.

\emph{Stochastic dynamics and Boltzmann distribution}. - We can model
the stochastic dynamics of this driven nonlinear system within our
amplitude equation approach. The heat bath responsible for the intrinsic
damping of the cantilever also produces a fluctuating force, which
can be added to the rhs of Eq. (\ref{Xeq}) as a white-noise term
$\xi_{X}(t)$. The fluctuation-dissipation-theorem demands $\left\langle \xi_{x}(t)\xi_{x}(0)\right\rangle =2\Gamma\tilde{T}\delta(t)$,
with a reduced temperature $\tilde{T}=k_{B}T/(m\delta x^{2}\gamma^{2})$.
This gives rise to a fluctuating power $\xi_{E}=\dot{x}\xi_{x}$ in
Eq. (\ref{dEdt}), which in turn leads to a stochastic force $\xi_{A}(t)=\xi_{E}(t)/(A\omega_{0}^{2})$
on the rhs of the equation of motion (\ref{dAdt}) for $A$, with
$\left\langle \xi_{A}(t)\xi_{A}(0)\right\rangle =(\Gamma\tilde{T}/\omega_{0}^{2})\,\delta(t)$.
The cantilever will thus settle into a dynamical, driven equilibrium
described by the Boltzmann distribution (Fig. \ref{cap:BoltzmannDist})

\begin{equation}
w(A)\propto A\exp\left[-\frac{U(A)}{\tilde{T}}\right]\,.\label{boltzmann}\end{equation}
The prefactor $A$ accounts for the phase space associated with the
two quadratures of motion. Formally, it is produced by an additional
deterministic term $\Gamma\tilde{T}/(2\omega_{0}^{2}A)$ on the rhs
of Eq. (\ref{dAdt}), which stems from rewriting the Langevin-version
of (\ref{dEdt}) in Ito-form. Comparison with a Langevin-Runge-Kutta
\cite{2004_Wilkie_NumericsSDE} simulation of the original Langevin
equations for $x$ and $\alpha$ shows good agreement. Shot noise
may be neglected for the high photon numbers of interest here ($n_{{\rm max}}\sim10^{6}$). 

As the scale $\Delta U$ of the corrugations in the oscillatory {}``washboard-potential''
$U(A)$ is set by the ratio $\mathcal{P}/\Gamma$, see Eq. (\ref{UA}),
it can become much larger than the light-induced contribution to the
static potential $V_{{\rm eff}}$, Eq. (\ref{Veff}), which scales
like $\mathcal{P}$. Thus, at small intrinsic damping $\Gamma$, the
dynamically unstable cantilever may display a strong confinement of
the oscillation amplitude $A$ in the wells of the dynamical potential,
even while radiation effects on the static Boltzmann distribution
$\exp[-V_{{\rm eff}}(x)/\tilde{T}]$ would be negligible. In our numerical
example, the scale $\Delta U$ corresponds to a real temperature in
the tens of Kelvin, below which the Boltzmann distribution clearly
shows the structure of multiple dynamical attractors. The Kramers
rate for thermal hopping between wells is $\tau_{{\rm th}}^{-1}=(\Gamma\Delta U/2\pi\omega_{0}^{4})\exp[-\Delta U/\tilde{T}]$.
Both at $x_{0}<0$ and large $x_{0}>0$ the global minimum of $U(A)$
is at $A=0$, and $w(A)$ returns to the distribution of the unperturbed
oscillator. 

\emph{Photothermal (bolometric) forces} can be another source of coupling
between light field and cantilever and were used for cooling a cantilever
\cite{2003_08_Vogel_PhotothermalForceOnCantilever,2004_12_HoehbergerKarrai_CoolingMicroleverNature}.
Due to the finite heat conductance of the cantilever, they feature
an additional time-lag $\tau$ between light intensity and force,
in addition to the time-lag between cantilever motion and intensity
already present in the case of radiation pressure. Such a force enters
the rhs of the cantilever equation of motion (\ref{Xeq}) in the form
$\mathcal{P}'\tau^{-1}\int_{-\infty}^{t}dt'\,|\alpha|^{2}(t')\exp[-(t-t')/\tau]$,
where we have considered the special case $\tau=0$ up to now. The
prefactor $\mathcal{P}'$ is equal to $\mathcal{P}$ multiplied by
the ratio between bolometric and radiation pressure force at constant
light intensity. The whole preceding analysis still applies, provided
we divide the sum in Eq. (\ref{PtildeRad}) by the factor $1+i\omega_{0}\tau$
and add up all contributions to $P_{{\rm rad}}$ and $U(A)$. The
oscillatory structure of the potential is fully retained. However,
for large $\omega_{0}\tau$, the total magnitude of this bolometric
contribution is suppressed, such that only the radiation pressure
effects remain. 

\emph{Hysteresis and 'latching' measurements}. - The presence of multiple
locally stable dynamical attractors naturally leads to hysteresis
upon sweeping an external parameter. In addition, the sensitivity
to parameter perturbations is greatly enhanced near transitions between
attractors. This could lend itself to the convenient measurement of
very small displacements (i.e. small perturbations $\delta x_{0}$,
on the sub-picometer scale): Sweeping back and forth in $x_{0}$ can
leave the cantilever in either of two attractors, depending on whether
$\delta x_{0}$ was large enough. Afterwards, the resulting (stable)
amplitude may be measured. A similar `latching' scheme has been proposed
and implemented in a very different context, using the dynamical bistability
of a strongly driven Josephson junction \cite{2005_01_Siddiqi_JBA_Basics}. 

\emph{Conclusions}. - We have studied the nonlinear classical dynamics
of high-finesse micromechanical optical cavities and discovered a
form of dynamical multistability that can arise in high-finesse cavities
with a large mechanical quality factor of the oscillating mirror.
The cantilever undergoes self-sustained oscillations whose amplitude
settles into one of several attractors, which we have mapped out in
parameter space. We have derived an analytical description of the
slow amplitude dynamics, introduced an effective dynamical potential,
incorporated the effects of thermal noise leading to tunneling between
the potential wells, and extended our analysis to photothermal forces.
Finally, we have pointed out that the intrinsic hysteresis may be
useful to implement a `latching' scheme for detecting small displacements.
The effects described here should be measurable in cavities that are
within the reach of current technology.

\emph{Acknowledgments}. - We thank J. Gambetta for useful hints on
stochastic simulations. F. M. acknowledges support by a DFG scholarship.
S. M. G. acknowledges support by NSF-ITR 0325580 and NSF-DMR 0342157.

\end{document}